\documentclass[%
 reprint,
superscriptaddress,
showkeys,
 amsmath,amssymb,
 aps,
pre, 
longbibliography,
]{revtex4-2}

\usepackage{graphicx}
\usepackage[caption=false]{subfig} 
\usepackage{mathtools}

\usepackage{float}
\usepackage{mathrsfs}
\usepackage{dcolumn}
\usepackage{bm}

\usepackage[dvipsnames]{xcolor}

\newcommand{\ee}{\end{equation}} 
\newcommand{\be}{\begin{equation}}

\usepackage[mathscr]{euscript}  

\usepackage{dutchcal}

\usepackage{xcolor}  

\usepackage{tcolorbox}
\usepackage{comment}
\usepackage{float}

\makeatletter
\newsavebox{\@brx}
\newcommand{\llangle}[1][]{\savebox{\@brx}{\(\m@th{#1\langle}\)}%
  \mathopen{\copy\@brx\kern-0.5\wd\@brx\usebox{\@brx}}}
\newcommand{\rrangle}[1][]{\savebox{\@brx}{\(\m@th{#1\rangle}\)}%
  \mathclose{\copy\@brx\kern-0.5\wd\@brx\usebox{\@brx}}}
\makeatother

\begin{document} 

\preprint{APS/123-QED}

\title{Information-fluctuation inequalities for collective response}

\author{Kristian St\o{}levik Olsen}
\thanks{\textit{Correspondence: kristian.olsen@hhu.de}}
\affiliation{Institute for Theoretical Physics II - Soft Matter, Heinrich Heine University Düsseldorf, D-40225 Düsseldorf, Germany}

\begin{abstract}
Hidden stochastic effects acting uniformly on a many‑particle system can generate strong correlations and macroscopic relative fluctuations that persist at large system sizes, even when the particles themselves remain causally independent. Here we derive a universal upper bound on relative fluctuations for a large class of observables, formulated in terms of a generalized mutual information between observable states and the hidden variable. This information–fluctuation inequality provides general insights into the principles governing collective response induced by global disorder. We demonstrate the result with applications to non‑interacting Brownian gases exposed to various types of dynamical disorder.
\end{abstract}

\maketitle

\textit{Introduction} ---  Many-body systems often exhibit striking properties that arise not from direct interactions but from hidden variables or external environmental processes. When distinct degrees of freedom respond to a common fluctuating driver, they can display strong correlations despite being causally disconnected. Such scenarios are naturally captured by the structure of conditionally independent stochastic variables \cite{simon1954spurious,reichenbach1991direction}. This structure is realized in a broad range of fluctuating environments, including gases of Brownian particles with collective resets \cite{biroli2023extreme,biroli2024exact,biroli2023critical}, systems with diffusing diffusivity \cite{mesquita2025dynamically_new}, and particles confined by fluctuating traps \cite{sabhapandit2024noninteracting,biroli2024dynamically}. Related effects also govern oscillator ensembles and flocking systems, where shared responses to disorder or obstacles generate coherent collective motion \cite{teramae2004robustness,hong2016phase}.
Despite intense recent activity, spanning theoretical advances \cite{boyer2025emerging,de2026dynamically,kulkarni2025dynamically,demauro2026tuningstrengthemergentcorrelations} as well as recent experimental realizations \cite{biroli2025experimental}, the universal principles governing the physics of such systems remain largely unexplored.

Because all correlations in conditionally independent ensembles stem from fluctuations of the hidden variable, their large-$n$ behavior raises subtle questions. In conventional many-body systems, even in the presence of local interactions, macroscopic observables typically become self-averaging, whereby relative fluctuations decay with system size $n$ \cite{wiseman1998finite}. For conditionally independent systems, however, the shared driving typically makes these systems deviate from this paradigm. Understanding the behavior of macroscopic fluctuations and the physical consequences in such systems therefore constitutes a natural and largely unexplored problem.

We demonstrate that in systems undergoing such collective response, fluctuations of macroscopic observables generically exhibit super-extensive scaling that breaks the self-averaging property. We show that relative fluctuations are universally upper-bounded by a generalized mutual information between the system states and the hidden variable, providing general insights into how macroscopic fluctuations behave in the presence of collective response.

\textit{Fluctuations of additive observables} --- Many physical observables of interest are additive, i.e., built as sums of single-component or local contributions. We consider any observable of the general additive form
\begin{equation}\label{eq:obs}
    \mathcal{O}_n(x_1,...,x_n) = \sum_{j=1}^n \mathcal{O}(x_j),
\end{equation}
defined by an arbitrary observable $\mathcal{O}(x)$. For future reference, we will refer to the observables $\mathcal{O}_n$ as \emph{macroscopic} and the individual contributions $\mathcal{O}$ as \emph{microscopic}.  We denote by $\Phi$ a hidden random variable that couples to the observable states, and denote by lowercase $\varphi$ possible realizations of this variable. The mean of Eq. (\ref{eq:obs}), conditioned on a fixed value of the hidden variable, is simply given by $\langle \mathcal{O}_n|\varphi \rangle  = n\langle \mathcal{O}|\varphi \rangle $ due to the conditional independence property. Averaging over the hidden variable $\varphi$ one has the expected extensive behavior $ {\langle \mathcal{O}_n\rangle} = n{\langle \mathcal{O}\rangle}$.  The variance of such observables can be written in terms of the covariance of the microscopic contributions
\begin{equation}
    \text{Var}(\mathcal{O}_n) = \sum_{j=1}^n\text{Var}(\mathcal{O}(x_j))  + \sum_{j\neq k} \text{Cov}(\mathcal{O}(x_j),\mathcal{O}(x_k)).
\end{equation}
If no inter-particle correlations were present, the covariance would vanish, leading to a simple extensive variance proportional to the single-particle variance. For shared coupling to $\varphi$, however, the law of total covariance tells us that 
\begin{equation}\label{eq:inter1}
    \text{Var}(\mathcal{O}_n) = n \text{Var}(\mathcal{O})  + n(n-1) \text{Var}_\varphi (\langle \mathcal{O}|\varphi \rangle),
\end{equation}
where $ \text{Var}_\varphi$ denotes the variance with respect to the hidden variable.
See the appendix for details leading to this expression. At large particle numbers, there is in general a quadratic super-extensive scaling of the variance, a direct result of the shared response to the hidden variable. We will use the coefficient of variation $ {c}_V^2(\mathcal{O}_n) =  \frac{ \text{Var}(\mathcal{O}_n) }{\langle \mathcal{O}_n\rangle^2}$ as a measure of relative size of fluctuations, which can be expressed as
\begin{equation}\label{eq:cvfirst}
    c_V^2(\mathcal{O}_n) = \frac{c_V^2(\mathcal{O}) }{n}    + \frac{(n-1)}{n} \frac{\text{Var}_\varphi (\langle \mathcal{O}|\varphi \rangle)}{\langle \mathcal{O}\rangle^2}.
\end{equation}
The first term on the right-hand-side is exactly the decay of relative fluctuations expected in non-interacting systems, while the second term stems from correlations. This term causes a remnant degree of fluctuations to persist even in the thermodynamic limit
\begin{equation}\label{eq:main}
   \lim_{n\to \infty}    {c}_V^2(\mathcal{O}_n)   =\frac{\text{Var}_\varphi (\langle \mathcal{O}|\varphi \rangle)}{\langle \mathcal{O}\rangle^2}.
\end{equation}
The lack of decaying relative fluctuations as one approaches the thermodynamic limit is often referred to as a lack of self-averaging. It is worth emphasizing that even systems with local interactions may display self-averaging behavior where the relative fluctuations vanish in the thermodynamic limit.  For example, in spatially extended lattice systems when the correlation length is smaller than the system size, the central limit theorem guarantees that coefficients of variation decay as $n^{-1/2}$ and vanish as the number of particles grows \cite{aharony1996absence}. Only in systems with long-range interactions or order, such as in critical phenomena, is super-extensive scaling expected \cite{aharony1996absence,wiseman1998finite}. Surprisingly, even though there are no direct interactions in our case, conditional independence is sufficient to destroy self-averaging. In fact, similarly to systems near critical points, conditional independence can be seen as an instance of infinite correlation length, as the hidden variable $\Phi$ affects all states independently of the topology or geometry of the state space.

\textit{Information-fluctuation relations} --- Since correlation-induced fluctuations can persist even in the thermodynamic limit, a relevant question is how large such remnant fluctuations can be. Information-theoretic methods provide a natural answer to this question.
Recently, ideas from information theory have lead to several novel results belonging to the family of fluctuation-response relations, where changes in observables are related to fluctuations and various information-theoretic or thermodynamic measures \cite{hasegawa2019uncertainty,dechant2020fluctuation,zheng2025universal,aslyamov2025nonequilibrium,bao2025nonlinearresponseidentitiesbounds,bao2024nonequilibrium}. Here we consider a simple version of such relations, valid in the non-perturbative regime, and providing insights into the fluctuations in Eqs. (\ref{eq:cvfirst}) and (\ref{eq:main}). Consider the $\chi_2$-distance $\chi_2(p \parallel q)$, which provides a measure of distance or divergence in the space of densities. It satisfies, in addition to its integral representation $\chi_2(p \parallel q) = \langle \left(p(x)/q(x)-1\right)^2\rangle_q$, a direct relation to the difference of a general observable $\mathcal{O}$ as one passes from a probability density $q(x)$ to $p(x)$, namely
\begin{equation} \label{eq:chi2}
 \frac{|\langle \mathcal{O}(x) \rangle_p-\langle \mathcal{O}(x) \rangle_q |^2}{\text{Var}_q (\mathcal{O})}  \leq  \chi_2(p \parallel q).
\end{equation}
A compact discussion of this statistical distance, its relations to other divergences, as well as connections to recent fluctuation-response relations can be found in the appendix. To establish a connection with Eq. (\ref{eq:cvfirst}), consider the case where one sets $p\to p(x|\varphi)$ and $q\to p(x)$. This gives an inequality that bounds the difference between the conditional expectation and the marginal
\begin{equation}
 \frac{|\langle \mathcal{O}|\varphi\rangle - \langle \mathcal{O}\rangle |^2}{\text{Var} (\mathcal{O})}  \leq \chi_2(p(\cdot|\varphi) \parallel p(\cdot)).
\end{equation}
This can be seen as a fluctuation-response relation, where the "response" in the mean of any observable to conditioning on a specific realization of $\varphi$ compared to the full ensemble is bounded by the ensemble fluctuations and a statistical distance. By averaging over $\varphi$ we see that $\langle |\langle \mathcal{O}|\varphi\rangle - \langle \mathcal{O} \rangle |^2\rangle_\varphi$ becomes the variance $\text{Var}_\varphi(\langle \mathcal{O}|\varphi \rangle)$. Hence we have
\begin{equation}\label{eq:intermed}
 \text{Var}_\varphi(\langle \mathcal{O}|\varphi \rangle)  \leq  \text{Var} (\mathcal{O})  \langle \chi_2(p(\cdot|\varphi) \parallel p(\cdot))\rangle_\varphi.
\end{equation}
Recall that the mutual information between general variables $X$ and $Y$ can be defined as $I(X;Y) = \langle D_\text{KL} (p_{X|Y}  \parallel p_X )\rangle_Y$, with $D_\text{KL}$ the Kullback-Leibler divergence $D_\text{KL}(p\parallel q) = \langle \log (p/q)\rangle_p$ \cite{cover1999elements}. The quantity appearing in Eq. (\ref{eq:intermed}) can be seen as a generalized mutual information where the $\chi_2$-distance is used instead.
We denote this as $I_{\chi_2}(X;Y) \equiv \langle \chi_2 (p_{X|Y}  \parallel p_X )\rangle_Y$.  
Combining this observation with Eq. (\ref{eq:cvfirst}) and Eq. (\ref{eq:intermed}), one obtains a finite-$n$ fluctuation-information inequality of the form
\begin{equation}\label{eq:mainfinite}
   \frac{ c_V(\mathcal{O}_n)}{c_V(\mathcal{O})} \leq \sqrt{\frac{1}{n} + \left(1 - \frac{1}{n}\right)  I_{\chi_2}(X;\Phi) }.
\end{equation}
This inequality constitutes the main result of this work, and demonstrates universal limits to fluctuations for conditionally independent processes. The right-hand-side only depends on the underlying laws of the states and the hidden variable, and is universal for any choice of observable. The $n^{-1/2}$ decay in Eq. (\ref{eq:mainfinite}) is exactly what is expected for self-averaging observables. Additionally, a constant system-size-independent term survives as $n\to \infty$, leading to
\begin{equation}\label{eq:main2}
   \lim_{n\to \infty} \frac{{c}_V(\mathcal{O}_n)}{{c}_V(\mathcal{O})} \leq  \sqrt{I_{\chi_2}(X;\Phi)},
\end{equation}
Intuitively, this bound states that the fluctuations that persist in the thermodynamic limit are determined not only by the fluctuations in the individual microscopic contributions to the observable, but also the degree to which the state variables and the hidden variable are coupled. Recall that the mutual information provides a measure of the reduction of uncertainty about the hidden variable when learning about the system state, or vice versa, and hence characterizes their statistical dependence. If no generalized mutual information exists between the states and the hidden variable, i.e., $I_{\chi_2}(X;\Phi)=0$, the fluctuations vanish. The generalized mutual information can be interpreted as exactly the quantity that connects the microscopic fluctuations ${c}_V(\mathcal{O})$ to the macroscopic level ${c}_V(\mathcal{O}_n)$, measuring the extent to which fluctuations can deviate from self-averaging behavior in the presence of the hidden variable. Finally, we note that if one instead of additive observables is interested in empirical averages of the form $\mathcal{A}_n = \frac{1}{n}\sum_{j=1}^n \mathcal{O}(x_j) = \mathcal{O}_n/n$, one may naturally express the ratio of coefficients of variations in Eq. (\ref{eq:mainfinite}) in terms of variances, namely $\frac{{c}_V(\mathcal{O}_n)}{{c}_V(\mathcal{O})}  = \sqrt{  \frac{\text{Var}(\mathcal{A}_n)}{\text{Var}(\mathcal{O})}}$. In terms of such observables, we therefore have
\begin{equation}
      \frac{\text{Var}(\mathcal{A}_n)}{\text{Var}(\mathcal{O})}\leq \frac{1}{n} + \left(1 - \frac{1}{n}\right)  I_{\chi_2}(X;\Phi).
\end{equation}
Below we verify and demonstrate these bounds with two applications. \\

\begin{figure}
    \centering
    \includegraphics[width=0.97\columnwidth]{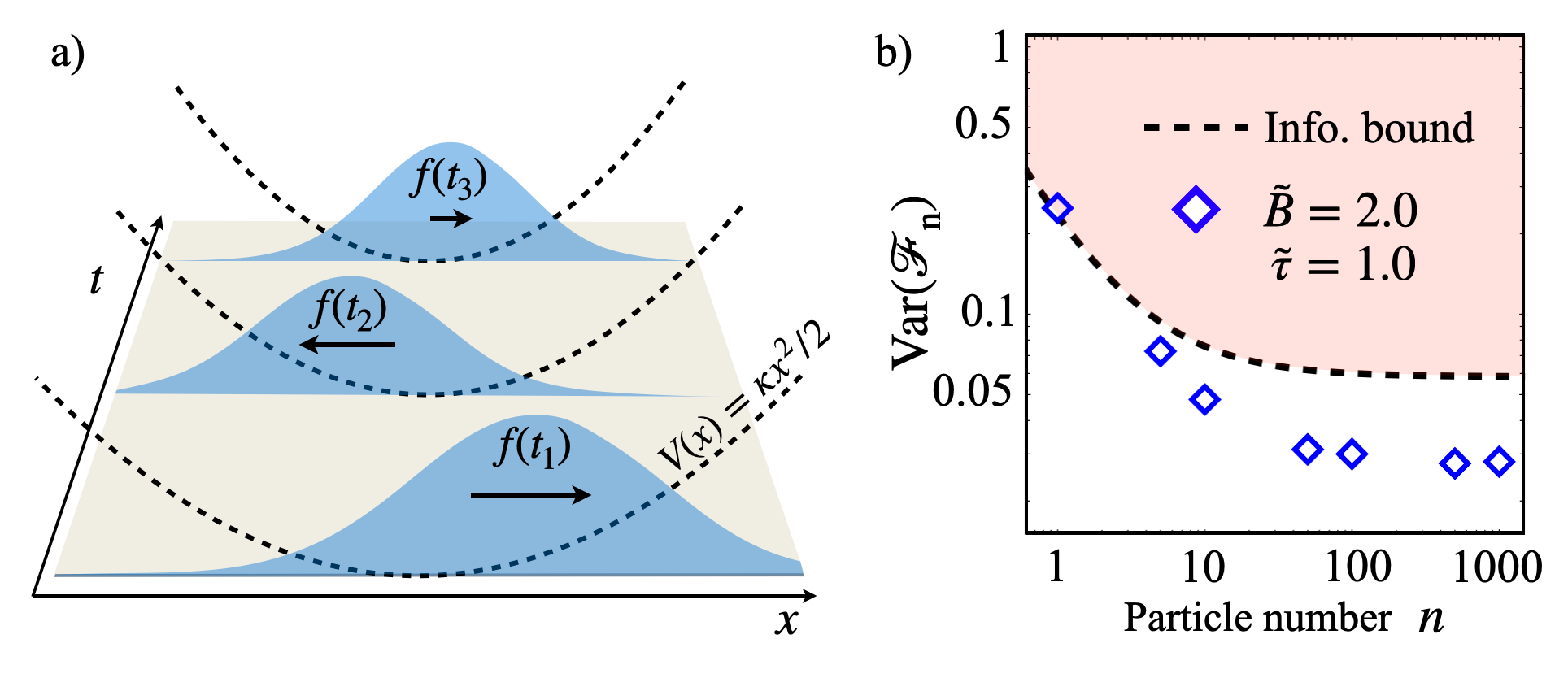}
     \caption{a) Sketch of the system, where the density of Brownian particles respond to a random external force. b) Variance of the number fluctuations near the origin as a function  of particle number. In panel b), dimensionless variables are used where lengths are set by $\sqrt{k_BT/\kappa}$, times by the relaxation timescale $\gamma/\kappa$, and characteristic force scale is set to $\sqrt{k_BT \kappa}$. This way only a dimensionless noise amplitude $\tilde B$ and timescale $\tilde \tau$ for the external Ornstein-Uhlenbeck forcing remains as independent parameters. }
    \label{fig:ou}
\end{figure}

\textit{Case I: Collective driving of a Brownian gas \& reaction rate fluctuations} --- Encounter‑based reaction rates play a central role in many physical, chemical and biological processes, from molecular reactions near reactive sites to cells interacting with binding regions \cite{grebenkov2020paradigm,bressloff2014stochastic,grebenkov2023diffusion}. In realistic environments, dynamical disorder is often unavoidable, which can strongly modulate the reaction rate \cite{zwanzig1990rate,doering1992resonant,lanoiselee2018diffusion}. We here bound the fluctuations of the instantaneous reaction rate in a reactive Brownian gas subject to a common stochastic driving force.

Consider a gas of $n$ Brownian particles confined to a stationary harmonic trap with stiffness $\kappa$. A stochastic force, for example originating from an externally applied field, drives the particles out of equilibrium (see Fig. (\ref{fig:ou}a)). The equation of motion for particle $j$ takes the form
\begin{equation}
    \gamma \dot x_j(t) = -\kappa x_j + f(t) + \sqrt{2 k_BT \gamma} \xi_j(t),
\end{equation}
where we take the external force to be an Ornstein-Uhlenbeck process $\dot f = -f/\tau + \sqrt{2B} \eta(t)$ with characteristic timescale $\tau$ and noise strength $B$, with a steady state distribution $P(f)$ that is Gaussian with zero mean and variance $B \tau$. The steady state of a particle conditioned on a value of the force is given by a Gaussian with variance $k_BT/\kappa$ and mean $f/\kappa$. The generalized mutual information can therefore be calculated from Gaussian integrals, resulting in $ I_{\chi_2}(X;F) = \frac{B \tau}{\kappa k_BT}$. For observables that are empirical averages this results in the bound
\begin{equation}
    \frac{\text{Var}(\mathcal{A}_n)}{ \text{Var}(\mathcal{O})} \leq  \frac{1}{n} + \left(1 -\frac{1}{n}\right)   { \frac{B \tau}{\kappa k_BT}}.
\end{equation}
This bounds the fluctuations of any observable of interest. In the context of reaction rates in the presence of dynamical disorder, we can relate reaction rate fluctuations to particle-number fluctuations. If each particle reacts at a rate $\lambda$ when residing in a spatial domain $\mathcal{D}\subset \mathbb{R}$, the instantaneous reaction rate is $r(t) = \lambda n \mathcal{F}_n(t)$, where $\mathcal{F}_n(t)$ is the stochastic fraction of particles in the reactive domain at any given time, given as an empirical mean of an indicator function
\begin{equation}
    \mathcal{F}_n = \frac{1}{n}\sum_{j=1}^n\mathbf{1}\{x_j(t)\in \mathcal{D}\}.
\end{equation}
For a single particle, this system is a Bernoulli process with probability $p_\mathcal{D} = \int_\mathcal{D} dx p(x)$, with $p(x)$ the single-particle steady-state density. The single-particle variance hence simply reads $\text{Var}(\mathcal{O}) = p_\mathcal{D}(1-p_\mathcal{D})$, giving rise to number fluctuations that are bounded by
\begin{equation}
    \text{Var}( \mathcal{F}_n)\leq\left[  \frac{1}{n} + \left(1 -\frac{1}{n}\right) { \frac{B \tau}{\kappa k_BT}} \right] p_\mathcal{D}(1-p_\mathcal{D}).
\end{equation}
In Fig. (\ref{fig:ou}b) we show using numerical simulations that the fluctuations are correctly bounded. In this case, we considered the central domain $\mathcal{D} = (-\ell,\ell)$, for which the single-particle residence probability is 
\begin{equation}
    p_\mathcal{D} = \text{erf}\left(\frac{\kappa  \ell}{\sqrt{2 \kappa   k_B T +2 B \tau }}\right).
\end{equation}
From the above bound one can directly get a bound for the fluctuations of the instantaneous reaction rate $r(t)$, namely $\text{Var}(r) = (\lambda n)^2\text{Var}(\mathcal{F}_n)$. In the thermodynamic limit we expect the bound
\begin{equation}
    \text{Var}\left( \frac{r}{n} \right)\leq   { \frac{B \tau\lambda^2 p_\mathcal{D}(1-p_\mathcal{D})}{\kappa k_BT}} 
\end{equation}
for the per-particle reaction rate. This bound allows one to relate the allowed extent of fluctuations to the properties of the dynamical disorder. Using the above dependence of $p_\mathcal{D}$ on $B$, one readily finds that the bound is linear at small noise strengths and scales as $\sim\sqrt{B}$ at large strengths. While the linear regime is expected from perturbation theory, the square-root scaling in the strongly disordered regime represents a non-perturbative bound where the responding fluctuations grow slower than those of the source.

\begin{figure}
    \centering
    \includegraphics[width=0.97\columnwidth]{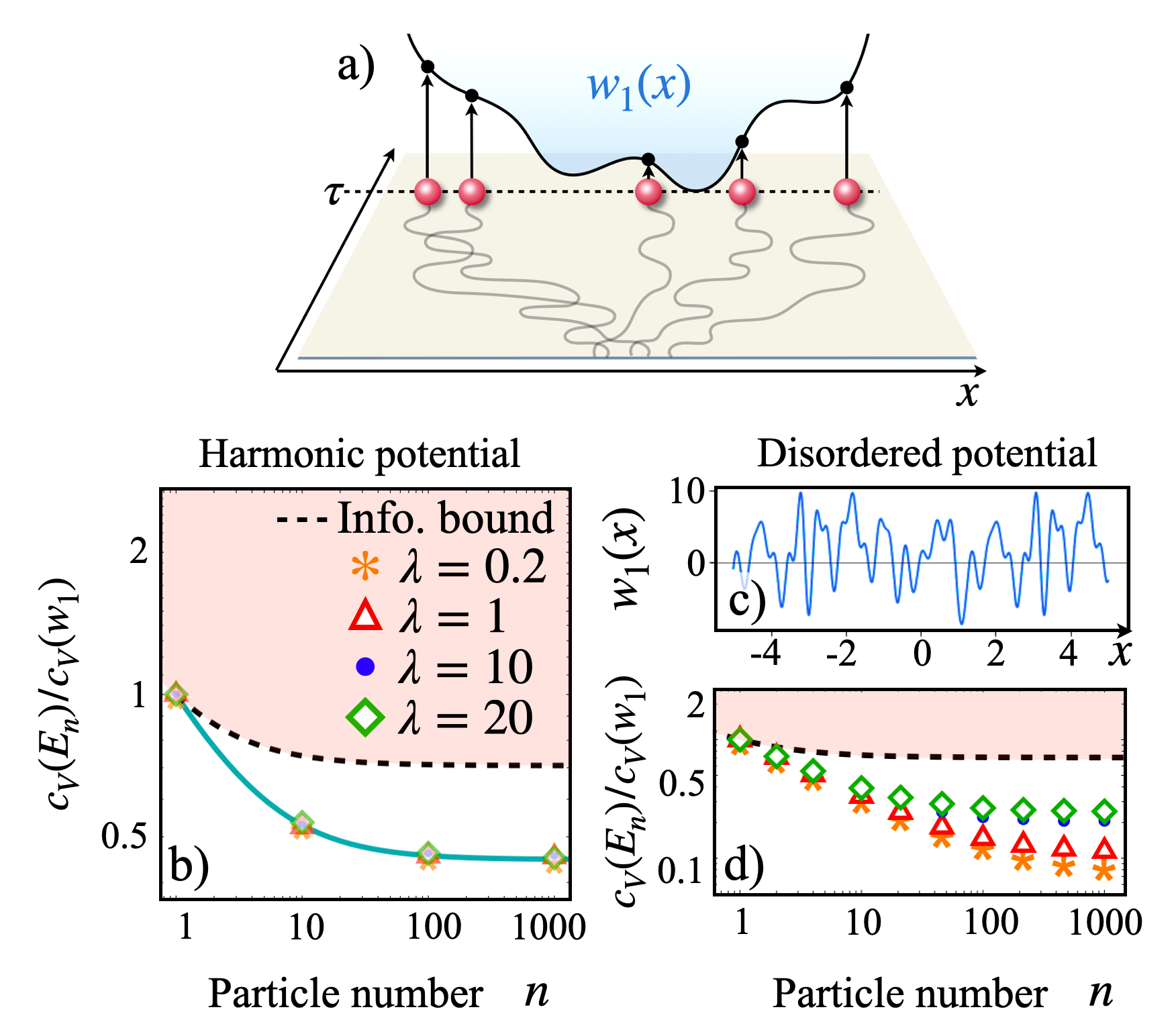}
     \caption{a) Independent Brownian particles evolve for a common time $\tau$, after which a potential is activated. Each particle gains an energy $w_1(x_j)$. We use exponentially distributed $\tau$ with rates $\lambda$ as indicated. b)  Comparison of the general finite-$n$ bound (dashed line) and Eq. (\ref{eq:trap_exact2}) for a harmonic potential. Dots show numerical simulations. c) Disordered potential used in panel d), showing numerical data from simulations using a disordered cosine potential with $20$ random contributions. Temporal units are chosen such that the trap relaxation time $\gamma/\kappa = 1$, and we set $D= 1/10$.}
    \label{fig:trap}
\end{figure}

\textit{Case II: Energetic cost of random activation of potentials} --- In recent years, much attention has been given to systems where a potential switches between two states in time. This generates non-equilibrium states, and is of relevance to models of random dynamical environments \cite{alston2022non,santra2021brownian} as well as stochastic resetting processes \cite{mercado2020intermittent,gupta2020stochastic,goerlich2025taming,olsen2024thermodynamic,gupta2025thermodynamic,olsen2024thermodynamic2}. Here we apply the above framework to a multi-particle system where a trap is randomly activated. As a relevant observable, we consider the energetic cost associated with the trap activation. Consider a period of time $\tau$ where Brownian particles diffuse freely, after which a harmonic trap is activated. The energetic cost of activating the trap in the presence of $n$ non-interacting particles is
\begin{equation}
    E_n(x_1,...,x_n) = \sum_{j=1}^n w_1(x_j), w_1(x_j) = \frac{\kappa}{2} x_j^2,
\end{equation}
where $ E_n$ denotes the total energetic cost, i.e., thermodynamic work, that must be payed, and $w_1$ the single-particle work (see Fig. (\ref{fig:trap}a) for an illustration). Conditioned on a fixed $\tau$, the mean energetic cost explicitly takes the form $\langle  E_n |\tau \rangle = \kappa nD \tau$, and the full mean $\langle  E_n  \rangle = \kappa nD \langle \tau\rangle$. Here we used the fact that the freely diffusing particles have Gaussian probability density with a variance equal to $2D\tau$ after time $\tau$. The variance in $ E_n$ conditioned on $\tau$ can also be calculated straightforwardly from Gaussian moment relations $\text{Var}( E_n|\tau)  =    2n  (\kappa D)^2  \tau^2   $. This leads to the following decomposition of the work fluctuations
\begin{equation}
       \text{Var}( E_n)  =   2n (\kappa D)^2 \langle \tau^2 \rangle +n^2 (\kappa D)^2 \text{Var}(\tau),
\end{equation}
displaying an extensive ($\sim n$) and super-extensive ($\sim n^2$) part as expected.
In terms of coefficients of variation we can express this as
\begin{equation}\label{eq:trap_exact2}
    \frac{c_V( E_n)}{c_V(w_1)}  = \sqrt{\frac{\left(1+ \frac{2}{n}\right) c_V^2(\tau)  +\frac{2}{n}}{ 3 c_V^2(\tau)  +2}},
\end{equation}
where we have also introduced the coefficient of variation of the hidden time variable. To compare this to the general bound, we consider the explicit case of an exponentially distributed time $\tau$, with density $\psi(\tau) = \lambda \exp(-\lambda \tau)$, for which $c_V(\tau) =1$. The generalized mutual information in this case involves not only Gaussian integrals  as in \textit{Case I}, but also Laplace distributions, and is shown in detail in the appendix. We find the simple result $I_{\chi_2}(X;\tau) = 1/2$, which leads to the bound
\begin{equation}\label{eq: FIBcaseI}
        \frac{c_V( E_n)}{c_V(w_1)}  \leq \sqrt{\frac{n+1}{2n}}.
\end{equation}
We compare this to Eq. (\ref{eq:trap_exact2}) with $c_V(\tau) = 1$, which reads $c_V( E_n)/ c_V(w_1)= \sqrt{(1+ \frac{4}{n})/5)}$. This comparison is shown in Fig. (\ref{fig:trap}b), showing clearly that the bound holds for all particle numbers $n$.

The above calculations could be performed exactly due to the linearity of the Langevin equation for a harmonic potential. Since the bound in Eq. (\ref{eq: FIBcaseI}) is universally valid for any observable, we verify it also for more complex potentials. We consider a disordered landscape $ w_1(x) = \sum_j a_j \cos(j x + b_j)$, where the random coefficients $a_j$ are taken to be normally distributed with zero mean and unit variance, and $b_j$ are uniform random phase shifts (see Fig. (\ref{fig:trap}c) for a specific realization). Fig. (\ref{fig:trap}d) shows an explicit comparison of the bound to numerical simulations where energy fluctuations were calculated. Since the energy $E_n$ of a single random activation of a potential landscape directly relates to the thermodynamic work and entropy needed to sustain an intermittent potential landscape over several activation cycles, these results may help constrain fluctuations of such thermodynamic quantities in the future.  \\

\textit{Discussion} --- We have shown that the mutual information between accessible system states and hidden variables governs fluctuation behavior in systems whose many degrees of freedom respond collectively to stochastic influences. In such settings, macroscopic observables exhibit super‑extensive scaling and obey a universal information–fluctuation bound expressed through a generalized mutual information. Two examples involving a Brownian gas subject to dynamical disorder illustrate these results: fluctuations of reaction rates in a Brownian gas  near a reactive patch, and work fluctuations arising when a potential landscape is activated at a shared random time, as in resetting processes. More broadly, the framework clarifies how large fluctuations can become in conditionally independent systems whose constituents respond coherently to shared external perturbations.

\acknowledgements
KSO acknowledges support from the Alexander von Humboldt foundation through their fellowship program. 

\appendix

\section{Variance formula derivation}
For the additive observables considered in Eq. (\ref{eq:obs}), we have a variance that is given by
\begin{equation}
    \text{Var}(\mathcal{O}_n) = \sum_{j=1}^n\text{Var}(\mathcal{O}(x_j))  + \sum_{j\neq k} \text{Cov}(\mathcal{O}(x_j),\mathcal{O}(x_k))
\end{equation}
We see that we must evaluate the covariances to proceed. The law of total covariance states that
\begin{align}
       \text{Cov}(\mathcal{O}(x_j),\mathcal{O}(x_k))  &=    \langle \text{Cov}( \mathcal{O}(x_j),\mathcal{O}(x_k)|\varphi)\rangle_\varphi \nonumber \\
       &+  \text{Cov}_\varphi (\langle \mathcal{O}(x_j)|\varphi\rangle,\langle \mathcal{O}(x_k)|\varphi\rangle) .
\end{align}
For the class of systems considered here, the first term vanishes from conditional independence, leaving only the covariance that arises between two realizations of the conditional mean as the hidden variable is varied collectively:
\begin{equation}
       \text{Cov}(\mathcal{O}(x_j),\mathcal{O}(x_k))  =    \text{Cov}_\varphi (\langle \mathcal{O}(x_j)|\varphi\rangle,\langle \mathcal{O}(x_k)|\varphi\rangle) .
\end{equation}
Since particles are identical, we have simply $\text{Cov}(\mathcal{O}(x_j),\mathcal{O}(x_k))  =     \text{Var}_\varphi (\langle \mathcal{O}(x_1)|\varphi \rangle)$, leading to a total variance
\begin{equation}\label{eq:inter1app}
    \text{Var}(\mathcal{O}_n) = n \text{Var}(\mathcal{O}(x_1))  + n(n-1) \text{Var}_\varphi (\langle \mathcal{O}(x_1)|\varphi \rangle).
\end{equation}
as was used in the main text.

\section{Bounds from the $\chi_2$-distance}
\begin{figure}
    \centering
    \includegraphics[width=7.5cm]{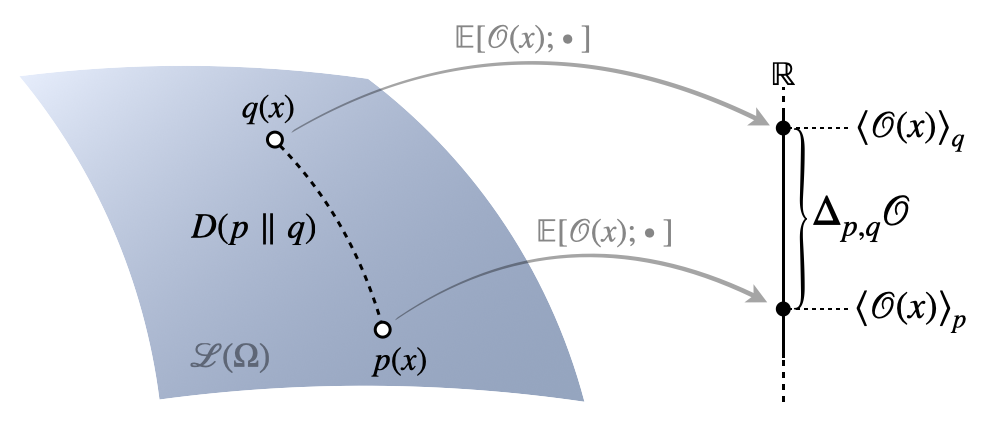}
    \caption{Under expectation values, a difference in probability densities as measured by generalized divergences $D(p\parallel q)$ on the space of densities $\mathcal{L}(\Omega)$ induces a change in an observable quantity $\Delta_{p,q}\mathcal{O}$. }
    \label{fig:dists}
\end{figure}

We let $\Omega$ denote the system's state space, i.e. $x\in \Omega$. Note that any expectation value with respect to a probability distribution with density $p(x)$, denoted $\mathbb{E}[\mathcal{O}(x);p(x)] \equiv \langle \mathcal{O}(x)\rangle_p$, can be interpreted as a map from the space of densities over state-space $\mathcal{L}(\Omega)$ to real numbers $ \mathbb{E}[\mathcal{O}(x);\cdot]: \mathcal{L}(\Omega)\to \mathbb{R}$. A central question of interest  in physics is what relations and bounds exist between the differences in observable outcomes $    \Delta_{p,q}\mathcal{O} \equiv \langle \mathcal{O}(x) \rangle_p-\langle \mathcal{O}(x) \rangle_q $ and distances in $\mathcal{L}(\Omega)$ (see Fig. \ref{fig:dists}).  While the difference in observable quantities are directly measurable, distances in the space of distributions belong to the realm of information geometry.  A natural statistical distance measure that relates directly to differences in observables is the $\chi_2$-distance, which in integral form is defined as
\begin{equation}
     \chi_2(p \parallel q) = \int_\Omega dx \frac{[p(x)-q(x)]^2}{q(x)}.
\end{equation}
The connection to how an observable changes as one passes from one probability measure to another is made clear through the variational definition of the $\chi_2$-distance, which reads \cite{polyanskiy2025information}
\begin{equation}
    \chi_2(p \parallel q) = \sup_{\mathcal{O}}\frac{ |\langle \mathcal{O}(x) \rangle_p-\langle \mathcal{O}(x) \rangle_q |^2 }{\text{Var}_q(\mathcal{O})},
\end{equation}
where the supremum is taken over observables.
Hence, the largest value the difference in expectation values can take for any admissible observable, relative to its variance, is given by the $\chi_2$-distance.
One therefore has the inequality
\begin{equation} \label{eq:chi2}
 \frac{|\langle \mathcal{O}(x) \rangle_p-\langle \mathcal{O}(x) \rangle_q |^2}{\text{Var}_q (\mathcal{O})}  \leq  \chi_2(p \parallel q),
\end{equation}
as was used in the main text.
This can also be obtained directly from the integral representation of the $\chi_2$-distance by using the Cauchy-Schwarz inequality. It is here worth noting the broader connection to  recent results in non-equilibrium statistical physics; considering a one-parameter family of densities $p_\theta(x)$, and choosing $q(x) =p_\theta(x) $ and $p(x) = p_{\theta+\Delta \theta}(x) $, we find to leading order $p_{\theta+\Delta \theta}(x) =p_{\theta}(x) + \Delta \theta \partial_\theta p_{\theta}(x) $ that $ |\Delta_{\theta+\Delta \theta,\theta}\mathcal{O} |^2  
\leq  \text{Var}_\theta (\mathcal{O})  \Delta \theta^2 \mathcal{I}(\theta)$ with $ \mathcal{I}(\theta)$ the Fisher information about $\theta$, given by $\mathcal{I}(\theta) = \langle (\partial_\theta\log p_\theta(x))^2\rangle_\theta$. Setting $\theta = 0$ results in a fluctuation-response relation, if we imagine that the parametric dependence on $\theta$ arises due to a perturbation of the underlying system with strength $\theta$ \cite{hasegawa2019uncertainty,dechant2020fluctuation}. Taylor expanding the left-hand side in $\Delta \theta$ results in $| \partial_\theta \langle \mathcal{O}(x) \rangle_{\theta}|^2 \leq \text{Var}_\theta (\mathcal{O})  \mathcal{I}(\theta) $. This bounds the rate at which a general observable can change when changing a parameter $\theta$. In the case of time-dependent probabilities, setting $\theta = t$ results in information-theoretic speed-limits \cite{ito2020stochastic}. While these results for parametric families of densities follow from Eq. (\ref{eq:chi2}), the inequality also holds for non-local densities that are not easily related by a change of a parameter, as is the case for the application in the main text.

\section{$\chi_2$-distance for hidden Poissonian waiting-time}
Consider the case of a non-interacting Brownian gas, left to freely diffuse for a random duration $\tau$. The single-article densities conditioned and marginalized over the time are given by
\begin{align}
    p(x|\tau) &= \frac{1}{\sqrt{4 \pi D \tau}} e^{-\frac{x^2}{4 D \tau}} \\
    p(x) &= \int_0^\infty d\tau \lambda e^{-\lambda \tau} p(x|\tau)\\
    & = \sqrt{\frac{\lambda}{4 D}} e^{-\sqrt{\frac{\lambda}{D}} |x|}
\end{align}
The generalized mutual information is calculated from the $\chi_2$-distance, which is this case can be calculated exactly as
\begin{align}
    \chi_2(p(x|\tau) \parallel p(x)) &= \int_{-\infty}^\infty dx \frac{[p(x|\tau) - p(x)]^2}{p(x)}\\
    & = \frac{e^{\frac{\lambda  \tau }{2}} }{ \sqrt{2 \pi \lambda  \tau }}  \left(2-\frac{\Gamma\left(\frac{1}{2},\frac{\lambda  \tau
   }{2}\right)}{\Gamma(1/2)}\right)-1
\end{align}
where $\Gamma(a,z)$ is  the incomplete Gamma function. To calculate the generalized mutual information we need $ I_{\chi_2}(x;\tau) = \langle  \chi_2(p(x|\tau) \parallel p(x))\rangle_\tau$. Once we average over $\tau$, the first term becomes
\begin{equation}
   2  \int_0^\infty d\tau \lambda e^{-\lambda \tau} \frac{e^{\frac{\lambda  \tau }{2}} }{ \sqrt{2 \pi \lambda  \tau }}  =    2 \int_0^\infty d u  \frac{e^{-u} }{ \sqrt{ \pi u }} 
\end{equation}
where we used the substitution $u = \lambda \tau/2$. The $u$-integral is nothing but the normalization of a gamma-distribution, and hence integrates to unity. Hence the integral we need to solve is
\begin{align}
     I_{\chi_2}(x;\tau) &=
    1 - \int_0^\infty d\tau \lambda e^{-\lambda \tau}  \frac{e^{\frac{\lambda  \tau }{2}} }{ \sqrt{2 \pi \lambda  \tau }}  \left(\frac{\Gamma\left(\frac{1}{2},\frac{\lambda  \tau
   }{2}\right)}{\Gamma(1/2)}\right)
\end{align}
Using the same substitution this can be expressed as
\begin{align}
     I_{\chi_2}(x;\tau) &=
    1 - \frac{1}{\pi}\int_0^\infty du \frac{e^{-u}}{\sqrt{u}} \Gamma(1/2,u)
\end{align}
The remaining $u$-integral can be written in a symmetric form using the integral definition of the incomplete  gamma function as
\begin{equation}
   \frac{1}{\pi} \int_0^\infty du \frac{e^{-u}}{\sqrt{u}} \Gamma(1/2,u) = \int_0^\infty du \frac{e^{-u}}{\sqrt{\pi u}} \int_u^\infty ds \frac{e^{-s}}{\sqrt{\pi s}}
\end{equation}
Since the integrand is symmetric in $u$ and $s$ and the integral is over half the $(u,s)$-space, we can write the integral simply as
\begin{equation}
   \frac{1}{\pi} \int_0^\infty du \frac{e^{-u}}{\sqrt{u}} \Gamma(1/2,u) = \frac{1}{2}\int_0^\infty du \frac{e^{-u}}{\sqrt{\pi u}} \int_0^\infty ds \frac{e^{-s}}{\sqrt{\pi s}}
\end{equation}
Again, the integrands in $u$ and $s$ are gamma distributions, hence by normalization the whole double integral integrates to $1/2$. Remarkably, the generalized mutual information therefore simply reads
\begin{equation}
    I_{\chi_2}(x;\tau) = \langle  \chi_2(p(x|\tau) \parallel p(x))\rangle_\tau =  \frac{1}{2}
\end{equation}
which results in the general bound in Eq. (\ref{eq: FIBcaseI}).

\section{Local equivalence of bounds}
The paper's main results, Eqs.(\ref{eq:mainfinite}-\ref{eq:main2}), are stated in terms of a generalized mutual information based on the $\chi_2$-distance. In cases where the conditional and marginal probability densities are sufficiently similar, the mutual information is expected to be small, and approximations can be made. In this regime, one can make use of a local equivalence between statistical distances to produce an equivalent family of bounds. 
This can be clearly seen by considering general $f$-divergences, defined by $D_f(p \parallel q) = \langle f(p(x)/q(x))\rangle_q$, where a function $f(t)$ satisfying $f(1)=0$ defines the divergence \cite{polyanskiy2025information,cover1999elements}. Consider a case where the distributions in question are pointwise close, with $p(x)/q(x) = 1 + \varepsilon(x)$. By Taylor expanding $ f(1+\varepsilon(x))$ to second order in $\varepsilon(x)$,
making use of $f(1) = 0$ and that $\langle \varepsilon(x) \rangle_q = 0$ by normalization, one finds $ D_f(p \parallel q) \approx \frac{f''(1)}{2} \langle \varepsilon^2(x) \rangle_q$. The $\chi_2$-distance is one example of a $f$-divergence obtained by picking $f(t) = |t-1|^2$. In this case $f''(1) = 2$ and $ \chi_2(p \parallel q) = \langle \varepsilon^2(x) \rangle_q $, leading to $    D_f(p \parallel q) \approx  \frac{f''(1)}{2} \chi_2(p \parallel q)$. Depending on the choice of the function $f(t)$, a simple rescaling $\frac{f''(1)}{2}$ allows one to pass between different statistical distance measures. This implies for the generalized mutual information in Eqs.(\ref{eq:mainfinite}-\ref{eq:main2}) the local relation
\begin{equation}
 I_{\chi_2}(X;Y)  \approx \frac{2}{f''(1)}   \langle D_f (p_{X|Y}  \parallel p_X ) \rangle_Y ,
\end{equation}
and we can  use a generalized mutual information based on any $f$-divergence if preferable. In particular, for $f(t) = t \log t$ with $f''(1) = 1$ one recovers the Kullback-Leibler divergence, and hence the normal mutual information. The fluctuation-information bound then reads
\begin{equation}\label{eq:main3}
   \lim_{n\to \infty} \frac{{c}_V(\mathcal{O}_n)}{{c}_V(\mathcal{O})} \leq  \sqrt{2 I(X;\Phi)},
\end{equation}
with the bound expressed in terms of the conventional mutual information. This bound only holds locally however, in cases where the degree of mixing or scrambling in the conditionally independent probability density is not too large. \\

To verify the local version of the bound given by Eq. (\ref{eq:main3}), consider again the example with Brownian gas diffusing for a random duration before a potential is activated (Case II). We consider the regime where the duration of free diffusion is only slightly non-deterministic, where we expect the marginal and conditional probabilities to be similar. The mutual information $ I(X;\tau) = H[X] - H[X|\tau]$ can be calculated by noting that $ H[X|\tau] = \frac{1}{2} \langle[1+ \log (4\pi D \tau)]\rangle_\tau$. For small fluctuations around the mean, we may expand this as
\begin{equation}
    H[X|\tau]  =  \frac{1}{2} + \frac{1}{2}  \log (4\pi D \langle \tau\rangle) - \frac{1}{4}\frac{\sigma_\tau^2}{\langle \tau \rangle^2} +O(\sigma_\tau^3)
\end{equation}
For small fluctuations in $\tau$, the marginal distribution  $p(x) = \langle p(x|\tau)\rangle_\tau$  can be obtained by expanding $p(x|\tau)$ in $\tau$ around the mean, leading to $p(x) \approx p(x|\langle\tau\rangle) + g(x)\sigma_\tau^2$. Since the variance of the full marginal is $2 D \langle \tau \rangle$, which exactly matches the variance of $p(x|\langle\tau\rangle) $, we have $\int dx g(x) x^2=0$. By normalization, $\int dx g(x) =0$. Using these facts, we may expand the entropy in small fluctuations $\sigma_\tau^2$, revealing that the entropy has a leading order correction that is fourth order in the standard deviations
\begin{equation}
    H[X] =  \frac{1}{2} + \frac{1}{2}  \log (4\pi D \langle \tau\rangle)  + O(\sigma_\tau^4)
\end{equation}
Hence the leading order behavior of the mutual information under small temporal fluctuations  is 
\begin{equation}
     I(X;\tau) \approx \frac{1}{4}\frac{\text{Var}(\tau)}{\langle \tau \rangle^2}  = \frac{c_V(\tau)}{4}
\end{equation}
The relative fluctuations of the work in the thermodynamic limit, obtained by letting $n\to \infty$ in Eq. (\ref{eq:trap_exact2}), can therefore be expressed in terms of the mutual information in this small-fluctuation regime:
\begin{equation}
    \frac{c_V(E_n)}{c_V(w_1)}  = \sqrt{\frac{2I(X;\tau) }{6 I(X;\tau) +1}}
\end{equation}
Clearly  the inequality $2 z/(6z+1)\leq 2z$ holds for any positive number $z$, and hence 
\begin{equation}
    \frac{c_V(E_n)}{c_V(w_1)}  = \sqrt{\frac{2I(X;\tau) }{6 I(X;\tau) +1}}
  \leq \sqrt{2I(X;\tau)}
\end{equation}
meaning the local bound in Eq. (\ref{eq:main3}) is indeed also satisfied.


%

\end{document}